\documentclass[sigchi-a]{acmart}
\usepackage{enumitem}

\AtBeginDocument{%
  \providecommand\BibTeX{{%
    \normalfont B\kern-0.5em{\scshape i\kern-0.25em b}\kern-0.8em\TeX}}}

\setcopyright{rightsretained}
\copyrightyear{2020}
\acmYear{2020}
\acmDOI{10.1145/1122445.1122456}

\acmConference[CHI '20]{}{April 26, 2020}{Honolulu, HI}
\acmBooktitle{CHI '20: ACM CHI Conference on Human Factors in Computing Systems,
  June 03--05, 2018, Woodstock, NY}
\acmPrice{15.00}
\acmISBN{978-1-4503-XXXX-X/18/06}

\begin{document}

\title{Usable, Acceptable, Appropriable: Towards Practicable Privacy}


\author{Aakash Gautam}
\affiliation{%
  \institution{Virginia Tech}}
\email{aakashg@vt.edu}



\begin{abstract}
A majority of the work on digital privacy and security has focused on users from developed countries who account for only around 20\% of the global population.
Moreover, the privacy needs for population that is already marginalized and vulnerable differ from users who have privilege to access a greater social support system. 
We reflect on our experiences of introducing computers and the Internet to a group of sex-trafficking survivors in Nepal and highlight a few socio-political factors that have influenced the design space around digital privacy.
These factors include the population's limited digital and text literacy skills and the fear of stigma against trafficked persons widely prevalent in Nepali society. 
We underscore the need to widen our perspective by focusing on \textit{practicable} privacy, that is, privacy practices that are (1) usable, (2) acceptable, and (3) appropriable. 
\end{abstract}

\maketitle

\section{Introduction}







Over the past three years, we have been working with an anti-trafficking non-governmental organization (NGO) in Nepal and exploring prospects for sex-trafficking survivors living in a shelter home. 
Undertaking an asset-based approach \cite{kretzmann1993building, mathie2005driving},  we have worked with a group of survivors\footnote{The survivors addressed each other as ``sisters''. I addressed the survivors as ``sisters'' as well. To match this nomenclature, I shall henceforth call the group we worked with ``sister-survivors''.} to identify their existing strengths and seek ways to build upon it.
One such way has involved the development and introduction of technology, and with it, we have encountered concerns around privacy and security.

The sister-survivors' needs for privacy around technology are different from the dominant discourse in the West due to a myriad of reasons including: (1) the sister-survivors are vulnerable, most have limited digital and text literacy, and have fewer opportunities to learn or use technology, (2) the sister-survivors have limited private physical space in the shelter home and are likely to have similar limited space after they leave the shelter home, (3) sex-trafficking survivors face stigma in Nepali society resulting in many being shunned by family and friends, and (4) Nepal, in general, can be considered a collectivist society \cite{hofstede2005cultures} so following some of the commonly acceptable privacy practices that stem from more-individualistic Western mores may lead to ostracization. 

\begin{marginfigure}
  \centering
  \includegraphics[width=1.1\linewidth]{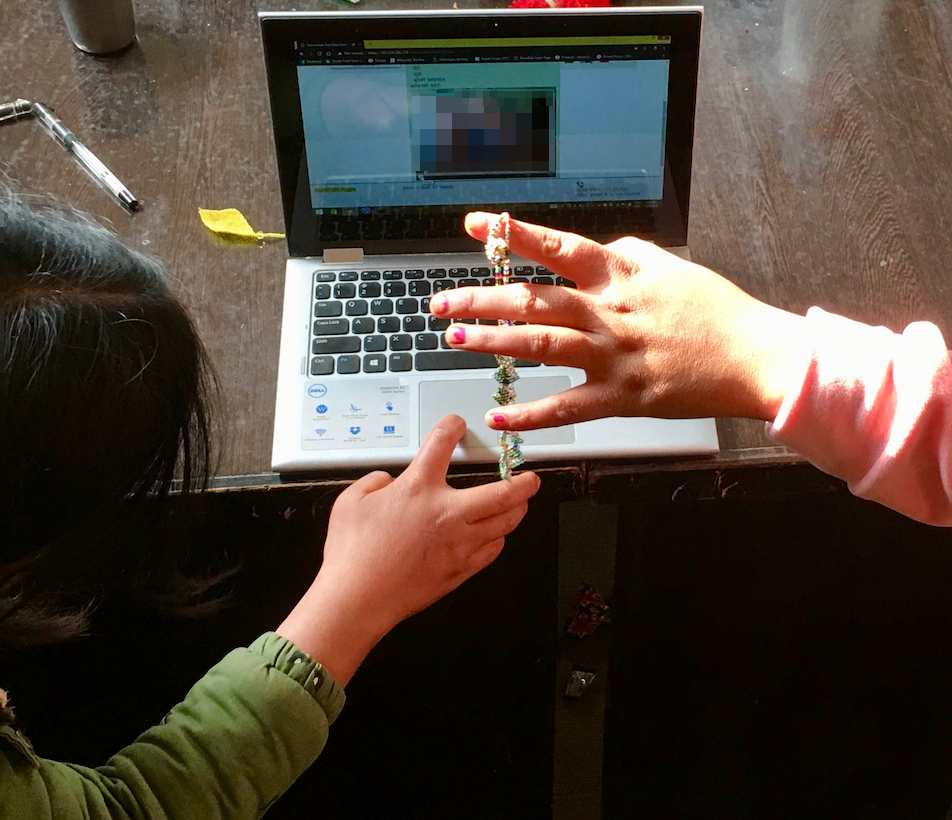}
  \caption{The sister-survivors feared being identified as a trafficked person. When we introduced a web application to a group of sister-survivors, we observed that they helped each other to hide when recording videos using a webcam. }
  \Description{When we introduced a web application to a group of sister-survivors, we observed that they helped each other to hide when recording videos using a webcam.}
  \label{fig:HamrokalaSupport}
\end{marginfigure}

One of the elements of strengths that the sister-survivors possessed is their strong mutual bond with one another \cite{gautam2018social}.
We observed the mutual bond being manifested in the wide range of support they provided each other when working on handicrafts, a task they did as part of their skill-based training program at the shelter home. 
When we introduced a web application to a group of sister-survivors, we noticed that they leveraged their mutual bond to use and appropriate the technology \cite{gautam2020chi}. 
They were also able to negotiate practices around the technology, including ways to protect their privacy (Figure \ref{fig:HamrokalaSupport}).
By leveraging their strengths, they were able to make the technology their own. 

Building on the experience, we believe that there is a possibility of leveraging the sister-survivors' mutual bond to support privacy practices around technology, even after they leave the shelter home. 
In particular, we see a possibility of presenting privacy, not as an individual's prerogative which is common in the West, but rather explore privacy as a collectively-held value requiring mutual support. 
We are currently exploring the communal use and appropriation of technology, and examining collective practices around privacy in the shelter home.
Overall, in this workshop, we hope to present some of our observations from the field,  the socio-cultural influences that shape our design space, learn about other participants' experiences, and chart out possible ways to support the sister-survivors' privacy and security around technology.

\section{Influential Factors}
In this section, we highlight some of the socio-cultural factors that have influenced our approach and have led us to seek alternative ways of thinking about privacy.

\subsection{Limited Digital and Text Literacy}

The sister-survivors typically have limited text and digital literacy \cite{gautam2018social, nhrc2017}. 
Privacy vulnerability arising due to lack of digital and text literacy has been documented both in developing \cite{vashistha2018examining, chen2014exploring} and developed countries \cite{olmstead2017americans, redmiles2017digital}.
Further, English has become the de facto language for many technology applications and lack of familiarity with it adds to the barrier in making informed digital privacy decisions.
Several web-based terminologies are phonetically translated in Nepali technologies with words such as ``login'' and ``logout'' being commonplace. 
However, these words held little meaning to the sister-survivors \cite{gautam2019voice}.

This reflects on practices around privacy as well. 
There is no Nepali word or phrase that encapsulates ideas of privacy, that is, of ownership and flow of data.
The closest words refer to secrecy, seclusion, or insulation.  
Thus, even before presenting potential technical approaches, we see a need for building common ground such as by engaging in participatory approaches so that the participants can define and act upon their own idea(s) of ownership and flow of data.

\marginpar{%
    \fbox{%
    \begin{minipage}{1.1\marginparwidth}
        Proposing a set of universal set privacy practices may not be helpful. Privacy practices have to be usable considering the varying skills possessed by the targeted group, culturally acceptable so that it does not lead to ostracization, and appropriable such that groups can adapt and make the practices their own.  We believe that the three conditions are critical to promoting practicable privacy, especially for those who are already in margins. 
        
        In our particular context, considering that the sister-survivors hold valuable strength in their mutual bond, these three practices could inform design in the following ways:
        \begin{itemize}[leftmargin=*]
            \item \textbf{Usable:} 
                \begin{itemize}[leftmargin=*]
                    \item Elicit and use terminologies that make sense to them; design to overcome digital and text literacy.
                    \item Design for collective use and support
                    \item Support obfuscation \cite{gautam2018social} to ensure they can have plausible deniability to protect themselves from stigma.
                \end{itemize}
            \item \textbf{Acceptable:} 
                \begin{itemize}[leftmargin=*]
                    \item Align privacy practice with the local culture such as the collectivist nature of society. 
                    \item Support multi-level group use and collaboration by making individual activities perceptible to others in the group. 
                \end{itemize}
            \item \textbf{Appropriable:}
                \begin{itemize}[leftmargin=*]
                    \item Design to ensure that, as a group, they can put their knowledge together to decide the next step.
                    \item Build support to make it easy to get started and encourage multiple ways of doing things.
                \end{itemize}
        \end{itemize}
    \end{minipage}}\label{threeList} }

\subsection{Collectivist Values and Lack of Power} 

Research already shows that people who are marginalized face great privacy vulnerability.
Studies conducted in the United States have shown how technology places the poor \cite{hess2017privacy, madden2017privacy, rainie2013anonymity}, elderly \cite{cornejo2016vulnerability}, and disabled people \cite{cornejo2016vulnerability} at risk to be scammed, subjected to fraud, stalked and impersonated. 
The interaction of two or more social disadvantages results in even greater digital vulnerability \cite{horrigan2016digital}. 
A contributing factor is that people with lower digital skills and socioeconomic standing (SES) have limited opportunity to learn about or seek support towards digital privacy \cite{redmiles2017digital}. 

The disadvantages discussed above are more pronounced when we look at privacy vulnerabilities in the Global South but so too are communal practices.  
Societies such as in Nepal, India and many countries in the Global South could be considered being more collectivist \cite{hofstede2005cultures}. 
Mobile phones are seen less as individually-owned information devices and more as shared communication devices. It raises several privacy-related concerns (e.g. \cite{ahmed2016privacy, sambasivan2018privacy, sambasivan2010intermediated}).
A significant part of information gathering in Nepal's context involves interactions with local people, friends, and families. 
Anecdotally, it is fairly common for either of my parents who are native Nepali, fairly well-educated and digitally literate, to hand over their phone to an employee in a mobile (repair) store and request help without encrypting or locking applications. 

Desiring for privacy could lead to being ostracized. 
The sister-survivors mentioned that doing something different than their family led them being ``othered'', with them being labeled as ``a haughty person who thinks she is better than us [family members]'' (S2). 
Fear from such ostracization is commonly held by the sister-survivors and is further accentuated by their fear of raising suspicion or being identified as a trafficked person.
The sister-survivors have limited power to negotiate such societal practices.
Seeking privacy, particularly around technology, is \emph{othered} by the sister-survivors. 
This othering resonates with Sambasivan \textit{et al.}'s study of women in South Asia who believed that privacy was ``for those rich women'' \cite{sambasivan2018privacy}.

\section{Our Prior Attempt and Next Step}

\subsection{Fear-Driven Practices}

Following the introduction of a web application connected to a local server, the sister-survivors expressed an interest in learning about and using the Internet. 
We discussed how the Internet works and the various ways in which we can use it to access information.
Following that, we asked what steps they took to keep themselves safe in their day-to-day life. 
In this group elicitation session, they mentioned practices such as ``do not talk to strangers'', ``do not go to unfamiliar places'', and ``keep an eye out for danger''.
We discussed and drew parallels to practices in the digital world. Synthesizing their day-to-day practices and extending it to the digital world led to a set of four easy-to-follow rules to remain safe while using the Internet (see sidebar). 

\marginpar{%
    \fbox{%
    \begin{minipage}{1.1\marginparwidth}
        The four rules that came up through the process were:
        \begin{enumerate}[leftmargin=*]
            \item Do not visit websites that you are not familiar with; ask the trainer about it before visiting
            \item Check for a padlock and ensure there is no red sign on the top of the browser before sharing information even for websites that you trust
            \item Check that there is no light near the webcam if you do not want to use a webcam
            \item Do not use Facebook or any website to contact people outside the shelter home
        \end{enumerate}

    \end{minipage}}\label{quote:Webcam} }


The rules were formulated ad hoc and they all suggest a fear-driven approach to privacy.
They do not convey positive values around the use of the Internet, and, more critically, could hinder adoption and appropriation moving forward.
In that sense, they are not \textit{practicable} in the long run. 

\subsection{Towards Collective Privacy Practices}

We noticed that the sister-survivors were supporting each other while using the Internet.
These include reminding others of the four rules, suggesting ways to navigate out of unsafe sites, and finding whether the website is safe or not. 
These lead us to believe that there is a possibility, in this context, of promoting privacy as a collective practice.  

Prior work has shown the value of social relationships in promoting learning of security and privacy practices (e.g. \cite{das2015role, rader2012stories, digioia2005social}). 
For example, Pierce \textit{et al.} \citeyear{pierce2018differential} found that most security tool were designed for individual users.
They posit that security and safety are socially contingent and hence there is a need for security and privacy that support collective action \cite{pierce2018differential}.

However, it is worth noting that collective action within a limited group may not be sufficient. 
This is especially true for people with limited digital literacy who may have fewer people in their network to clarify and support in technology-related concerns, and they may develop mental models that hinder their privacy decisions \cite{vitak2018knew, besnard2004mental}.
Thus we are exploring the possibility of multi-level groups \textit{defined by the user} -- friends and families, locally situated individuals, NGO staff members, and external experts and curated resources -- to support collective privacy practices.


\section{Conclusion: Towards Practicable Privacy}

The sister-survivors' limited digital and text literacy skills and their fear of being stigmatized in society being identified as a trafficked person, defines our design space. 
Further, we have to be cognizant of broader society's values and orientation. 
In our case, Nepali society is collectivist to a large extent and we observed similar orientation among the sister-survivors \cite{gautam2020chi}.

We hope to call for privacy practices that vulnerable populations can use without fear of social exclusion and can modify it to adjust it to their needs and values. 
This leads us to three critical conditions that we believe are required in any socio-technical systems to promote privacy:
(1) usable, (2) socially acceptable, and (3) appropriable. 
While these conditions are not exhaustive, we believe that they will help create practicable privacy for vulnerable populations. 






\bibliographystyle{ACM-Reference-Format}
\bibliography{sample-base}


\begin{thebibliography}{25}


\ifx \showCODEN    \undefined \def \showCODEN     #1{\unskip}     \fi
\ifx \showDOI      \undefined \def \showDOI       #1{#1}\fi
\ifx \showISBNx    \undefined \def \showISBNx     #1{\unskip}     \fi
\ifx \showISBNxiii \undefined \def \showISBNxiii  #1{\unskip}     \fi
\ifx \showISSN     \undefined \def \showISSN      #1{\unskip}     \fi
\ifx \showLCCN     \undefined \def \showLCCN      #1{\unskip}     \fi
\ifx \shownote     \undefined \def \shownote      #1{#1}          \fi
\ifx \showarticletitle \undefined \def \showarticletitle #1{#1}   \fi
\ifx \showURL      \undefined \def \showURL       {\relax}        \fi
\providecommand\bibfield[2]{#2}
\providecommand\bibinfo[2]{#2}
\providecommand\natexlab[1]{#1}
\providecommand\showeprint[2][]{arXiv:#2}

\bibitem[\protect\citeauthoryear{Ahmed, Guha, Rifat, Shezan, and Dell}{Ahmed
  et~al\mbox{.}}{2016}]%
        {ahmed2016privacy}
\bibfield{author}{\bibinfo{person}{Syed~Ishtiaque Ahmed},
  \bibinfo{person}{Shion Guha}, \bibinfo{person}{Md~Rashidujjaman Rifat},
  \bibinfo{person}{Faysal~Hossain Shezan}, {and} \bibinfo{person}{Nicola
  Dell}.} \bibinfo{year}{2016}\natexlab{}.
\newblock \showarticletitle{Privacy in repair: An analysis of the privacy
  challenges surrounding broken digital artifacts in bangladesh}. In
  \bibinfo{booktitle}{\emph{Proceedings of the Eighth International Conference
  on Information and Communication Technologies and Development}}. ACM,
  \bibinfo{pages}{11}.
\newblock


\bibitem[\protect\citeauthoryear{Besnard, Greathead, and Baxter}{Besnard
  et~al\mbox{.}}{2004}]%
        {besnard2004mental}
\bibfield{author}{\bibinfo{person}{Denis Besnard}, \bibinfo{person}{David
  Greathead}, {and} \bibinfo{person}{Gordon Baxter}.}
  \bibinfo{year}{2004}\natexlab{}.
\newblock \showarticletitle{When mental models go wrong: co-occurrences in
  dynamic, critical systems}.
\newblock \bibinfo{journal}{\emph{International Journal of Human-Computer
  Studies}} \bibinfo{volume}{60}, \bibinfo{number}{1} (\bibinfo{year}{2004}),
  \bibinfo{pages}{117--128}.
\newblock


\bibitem[\protect\citeauthoryear{Chen, Paik, and McCabe}{Chen
  et~al\mbox{.}}{2014}]%
        {chen2014exploring}
\bibfield{author}{\bibinfo{person}{Jay Chen}, \bibinfo{person}{Michael Paik},
  {and} \bibinfo{person}{Kelly McCabe}.} \bibinfo{year}{2014}\natexlab{}.
\newblock \showarticletitle{Exploring internet security perceptions and
  practices in urban ghana}. In \bibinfo{booktitle}{\emph{10th Symposium On
  Usable Privacy and Security ($\{$SOUPS$\}$ 2014)}}.
  \bibinfo{pages}{129--142}.
\newblock


\bibitem[\protect\citeauthoryear{Cornejo, Brewer, Edasis, and Piper}{Cornejo
  et~al\mbox{.}}{2016}]%
        {cornejo2016vulnerability}
\bibfield{author}{\bibinfo{person}{Raymundo Cornejo}, \bibinfo{person}{Robin
  Brewer}, \bibinfo{person}{Caroline Edasis}, {and} \bibinfo{person}{Anne~Marie
  Piper}.} \bibinfo{year}{2016}\natexlab{}.
\newblock \showarticletitle{Vulnerability, sharing, and privacy: Analyzing art
  therapy for older adults with dementia}. In
  \bibinfo{booktitle}{\emph{Proceedings of the 19th ACM Conference on
  Computer-Supported Cooperative Work \& Social Computing}}. ACM,
  \bibinfo{pages}{1572--1583}.
\newblock


\bibitem[\protect\citeauthoryear{Das, Kramer, Dabbish, and Hong}{Das
  et~al\mbox{.}}{2015}]%
        {das2015role}
\bibfield{author}{\bibinfo{person}{Sauvik Das}, \bibinfo{person}{Adam~DI
  Kramer}, \bibinfo{person}{Laura~A Dabbish}, {and} \bibinfo{person}{Jason~I
  Hong}.} \bibinfo{year}{2015}\natexlab{}.
\newblock \showarticletitle{The role of social influence in security feature
  adoption}. In \bibinfo{booktitle}{\emph{Proceedings of the 18th ACM
  conference on computer supported cooperative work \& social computing}}. ACM,
  \bibinfo{pages}{1416--1426}.
\newblock


\bibitem[\protect\citeauthoryear{DiGioia and Dourish}{DiGioia and
  Dourish}{2005}]%
        {digioia2005social}
\bibfield{author}{\bibinfo{person}{Paul DiGioia} {and} \bibinfo{person}{Paul
  Dourish}.} \bibinfo{year}{2005}\natexlab{}.
\newblock \showarticletitle{Social navigation as a model for usable security}.
  In \bibinfo{booktitle}{\emph{Proceedings of the 2005 symposium on Usable
  privacy and security}}. ACM, \bibinfo{pages}{101--108}.
\newblock


\bibitem[\protect\citeauthoryear{Gautam, Shrestha, Tatar, and Harrison}{Gautam
  et~al\mbox{.}}{2018}]%
        {gautam2018social}
\bibfield{author}{\bibinfo{person}{Aakash Gautam}, \bibinfo{person}{Chandani
  Shrestha}, \bibinfo{person}{Deborah Tatar}, {and} \bibinfo{person}{Steve
  Harrison}.} \bibinfo{year}{2018}\natexlab{}.
\newblock \showarticletitle{Social Photo-Elicitation: The Use of Communal
  Production of Meaning to Hear a Vulnerable Population}.
\newblock \bibinfo{journal}{\emph{Proceedings of the ACM on Human-Computer
  Interaction}} \bibinfo{volume}{2}, \bibinfo{number}{CSCW}
  (\bibinfo{year}{2018}), \bibinfo{pages}{56}.
\newblock


\bibitem[\protect\citeauthoryear{Gautam, Tatar, and Harrison}{Gautam
  et~al\mbox{.}}{2019}]%
        {gautam2019voice}
\bibfield{author}{\bibinfo{person}{Aakash Gautam}, \bibinfo{person}{Deborah
  Tatar}, {and} \bibinfo{person}{Steve Harrison}.}
  \bibinfo{year}{2019}\natexlab{}.
\newblock \showarticletitle{Adding Voices to Support Web Navigation Among a Low
  Digital Literacy Group}. In \bibinfo{booktitle}{\emph{Publication of the 2019
  on Designing Interactive Systems Conference 2019 Companion}}.
  \bibinfo{publisher}{ACM}, \bibinfo{address}{New York, NY, USA},
  \bibinfo{pages}{165--169}.
\newblock


\bibitem[\protect\citeauthoryear{Gautam, Tatar, and Harrison}{Gautam
  et~al\mbox{.}}{2020}]%
        {gautam2020chi}
\bibfield{author}{\bibinfo{person}{Aakash Gautam}, \bibinfo{person}{Deborah
  Tatar}, {and} \bibinfo{person}{Steve Harrison}.}
  \bibinfo{year}{2020}\natexlab{}.
\newblock \showarticletitle{Crafting, Communality, and Computing: Building on
  Existing Strengths To Support a Vulnerable Population}.
\newblock \bibinfo{journal}{\emph{Proceedings of the 2020 CHI Conference on
  Human Factors in Computing Systems}} (\bibinfo{year}{2020}).
\newblock


\bibitem[\protect\citeauthoryear{Hess}{Hess}{2017}]%
        {hess2017privacy}
\bibfield{author}{\bibinfo{person}{Amanda Hess}.}
  \bibinfo{year}{2017}\natexlab{}.
\newblock \showarticletitle{How privacy became a commodity for the rich and
  powerful}.
\newblock \bibinfo{journal}{\emph{The New York Times}}  \bibinfo{volume}{9}
  (\bibinfo{year}{2017}).
\newblock


\bibitem[\protect\citeauthoryear{Hofstede, Hofstede, and Minkov}{Hofstede
  et~al\mbox{.}}{2005}]%
        {hofstede2005cultures}
\bibfield{author}{\bibinfo{person}{Geert Hofstede}, \bibinfo{person}{Gert~Jan
  Hofstede}, {and} \bibinfo{person}{Michael Minkov}.}
  \bibinfo{year}{2005}\natexlab{}.
\newblock \bibinfo{booktitle}{\emph{Cultures and organizations: Software of the
  mind}}. Vol.~\bibinfo{volume}{2}.
\newblock \bibinfo{publisher}{Citeseer}.
\newblock


\bibitem[\protect\citeauthoryear{Horrigan}{Horrigan}{2016}]%
        {horrigan2016digital}
\bibfield{author}{\bibinfo{person}{John~B Horrigan}.}
  \bibinfo{year}{2016}\natexlab{}.
\newblock \showarticletitle{Digital Readiness Gaps.}
\newblock \bibinfo{journal}{\emph{Pew Research Center}} (\bibinfo{year}{2016}).
\newblock


\bibitem[\protect\citeauthoryear{Kretzmann and McKnight}{Kretzmann and
  McKnight}{1993}]%
        {kretzmann1993building}
\bibfield{author}{\bibinfo{person}{John~P Kretzmann} {and}
  \bibinfo{person}{John McKnight}.} \bibinfo{year}{1993}\natexlab{}.
\newblock \bibinfo{booktitle}{\emph{Building communities from the inside out}}.
\newblock \bibinfo{publisher}{Center for Urban Affairs and Policy Research,
  Neighborhood Innovations Network}.
\newblock


\bibitem[\protect\citeauthoryear{Kshetri, Banjade, and Subedi}{Kshetri
  et~al\mbox{.}}{2017}]%
        {nhrc2017}
\bibfield{author}{\bibinfo{person}{Kamal~Thapa Kshetri},
  \bibinfo{person}{Yesoda Banjade}, {and} \bibinfo{person}{Govind Subedi}.}
  \bibinfo{year}{2017}\natexlab{}.
\newblock \bibinfo{booktitle}{\emph{Trafficking in Persons National Report}}.
\newblock \bibinfo{type}{{T}echnical {R}eport}. \bibinfo{institution}{National
  Human Rights Commission Nepal}.
\newblock
\urldef\tempurl%
\url{http://www.nhrcnepal.org/nhrc_new/doc/newsletter/TIP_National_Report_2015_2016.pdf}
\showURL{%
\tempurl}


\bibitem[\protect\citeauthoryear{Madden, Gilman, Levy, and Marwick}{Madden
  et~al\mbox{.}}{2017}]%
        {madden2017privacy}
\bibfield{author}{\bibinfo{person}{Mary Madden}, \bibinfo{person}{Michele
  Gilman}, \bibinfo{person}{Karen Levy}, {and} \bibinfo{person}{Alice
  Marwick}.} \bibinfo{year}{2017}\natexlab{}.
\newblock \showarticletitle{Privacy, poverty, and big data: A matrix of
  vulnerabilities for poor Americans}.
\newblock \bibinfo{journal}{\emph{Wash. UL Rev.}}  \bibinfo{volume}{95}
  (\bibinfo{year}{2017}), \bibinfo{pages}{53}.
\newblock


\bibitem[\protect\citeauthoryear{Mathie and Cunningham}{Mathie and
  Cunningham}{2005}]%
        {mathie2005driving}
\bibfield{author}{\bibinfo{person}{Alison Mathie} {and} \bibinfo{person}{Gord
  Cunningham}.} \bibinfo{year}{2005}\natexlab{}.
\newblock \showarticletitle{Who is driving development? Reflections on the
  transformative potential of asset-based community development}.
\newblock \bibinfo{journal}{\emph{Canadian Journal of Development Studies/Revue
  canadienne d'{\'e}tudes du d{\'e}veloppement}} \bibinfo{volume}{26},
  \bibinfo{number}{1} (\bibinfo{year}{2005}), \bibinfo{pages}{175--186}.
\newblock


\bibitem[\protect\citeauthoryear{Olmstead and Smith}{Olmstead and
  Smith}{2017}]%
        {olmstead2017americans}
\bibfield{author}{\bibinfo{person}{Kenneth Olmstead} {and}
  \bibinfo{person}{Aaron Smith}.} \bibinfo{year}{2017}\natexlab{}.
\newblock \showarticletitle{Americans and cybersecurity}.
\newblock \bibinfo{journal}{\emph{Pew Research Center}}  \bibinfo{volume}{26}
  (\bibinfo{year}{2017}).
\newblock


\bibitem[\protect\citeauthoryear{Pierce, Fox, Merrill, and Wong}{Pierce
  et~al\mbox{.}}{2018}]%
        {pierce2018differential}
\bibfield{author}{\bibinfo{person}{James Pierce}, \bibinfo{person}{Sarah Fox},
  \bibinfo{person}{Nick Merrill}, {and} \bibinfo{person}{Richmond Wong}.}
  \bibinfo{year}{2018}\natexlab{}.
\newblock \showarticletitle{Differential Vulnerabilities and a Diversity of
  Tactics: What Toolkits Teach Us About Cybersecurity}.
\newblock \bibinfo{journal}{\emph{Proceedings of the ACM on Human-Computer
  Interaction}} \bibinfo{volume}{2}, \bibinfo{number}{CSCW}
  (\bibinfo{year}{2018}), \bibinfo{pages}{139}.
\newblock


\bibitem[\protect\citeauthoryear{Rader, Wash, and Brooks}{Rader
  et~al\mbox{.}}{2012}]%
        {rader2012stories}
\bibfield{author}{\bibinfo{person}{Emilee Rader}, \bibinfo{person}{Rick Wash},
  {and} \bibinfo{person}{Brandon Brooks}.} \bibinfo{year}{2012}\natexlab{}.
\newblock \showarticletitle{Stories as informal lessons about security}. In
  \bibinfo{booktitle}{\emph{Proceedings of the Eighth Symposium on Usable
  Privacy and Security}}. ACM, \bibinfo{pages}{6}.
\newblock


\bibitem[\protect\citeauthoryear{Rainie, Kiesler, Kang, Madden, Duggan, Brown,
  and Dabbish}{Rainie et~al\mbox{.}}{2013}]%
        {rainie2013anonymity}
\bibfield{author}{\bibinfo{person}{Lee Rainie}, \bibinfo{person}{Sara Kiesler},
  \bibinfo{person}{Ruogu Kang}, \bibinfo{person}{Mary Madden},
  \bibinfo{person}{Maeve Duggan}, \bibinfo{person}{Stephanie Brown}, {and}
  \bibinfo{person}{Laura Dabbish}.} \bibinfo{year}{2013}\natexlab{}.
\newblock \showarticletitle{Anonymity, privacy, and security online}.
\newblock \bibinfo{journal}{\emph{Pew Research Center}}  \bibinfo{volume}{5}
  (\bibinfo{year}{2013}).
\newblock


\bibitem[\protect\citeauthoryear{Redmiles, Kross, and Mazurek}{Redmiles
  et~al\mbox{.}}{2017}]%
        {redmiles2017digital}
\bibfield{author}{\bibinfo{person}{Elissa~M Redmiles}, \bibinfo{person}{Sean
  Kross}, {and} \bibinfo{person}{Michelle~L Mazurek}.}
  \bibinfo{year}{2017}\natexlab{}.
\newblock \showarticletitle{Where is the digital divide?: A survey of security,
  privacy, and socioeconomics}. In \bibinfo{booktitle}{\emph{Proceedings of the
  2017 CHI Conference on Human Factors in Computing Systems}}. ACM,
  \bibinfo{pages}{931--936}.
\newblock


\bibitem[\protect\citeauthoryear{Sambasivan, Checkley, Batool, Ahmed, Nemer,
  Gayt{\'a}n-Lugo, Matthews, Consolvo, and Churchill}{Sambasivan
  et~al\mbox{.}}{2018}]%
        {sambasivan2018privacy}
\bibfield{author}{\bibinfo{person}{Nithya Sambasivan}, \bibinfo{person}{Garen
  Checkley}, \bibinfo{person}{Amna Batool}, \bibinfo{person}{Nova Ahmed},
  \bibinfo{person}{David Nemer}, \bibinfo{person}{Laura~Sanely
  Gayt{\'a}n-Lugo}, \bibinfo{person}{Tara Matthews}, \bibinfo{person}{Sunny
  Consolvo}, {and} \bibinfo{person}{Elizabeth Churchill}.}
  \bibinfo{year}{2018}\natexlab{}.
\newblock \showarticletitle{``Privacy is not for me, it's for those rich
  women'': Performative Privacy Practices on Mobile Phones by Women in South
  Asia}. In \bibinfo{booktitle}{\emph{Fourteenth Symposium on Usable Privacy
  and Security ($\{$SOUPS$\}$ 2018)}}. \bibinfo{pages}{127--142}.
\newblock


\bibitem[\protect\citeauthoryear{Sambasivan, Cutrell, Toyama, and
  Nardi}{Sambasivan et~al\mbox{.}}{2010}]%
        {sambasivan2010intermediated}
\bibfield{author}{\bibinfo{person}{Nithya Sambasivan}, \bibinfo{person}{Ed
  Cutrell}, \bibinfo{person}{Kentaro Toyama}, {and} \bibinfo{person}{Bonnie
  Nardi}.} \bibinfo{year}{2010}\natexlab{}.
\newblock \showarticletitle{Intermediated technology use in developing
  communities}. In \bibinfo{booktitle}{\emph{Proceedings of the SIGCHI
  Conference on Human Factors in Computing Systems}}. ACM,
  \bibinfo{pages}{2583--2592}.
\newblock


\bibitem[\protect\citeauthoryear{Vashistha, Anderson, and Mare}{Vashistha
  et~al\mbox{.}}{2018}]%
        {vashistha2018examining}
\bibfield{author}{\bibinfo{person}{Aditya Vashistha}, \bibinfo{person}{Richard
  Anderson}, {and} \bibinfo{person}{Shrirang Mare}.}
  \bibinfo{year}{2018}\natexlab{}.
\newblock \showarticletitle{Examining security and privacy research in
  developing regions}. In \bibinfo{booktitle}{\emph{Proceedings of the 1st ACM
  SIGCAS Conference on Computing and Sustainable Societies}}. ACM,
  \bibinfo{pages}{25}.
\newblock


\bibitem[\protect\citeauthoryear{Vitak, Liao, Subramaniam, and Kumar}{Vitak
  et~al\mbox{.}}{2018}]%
        {vitak2018knew}
\bibfield{author}{\bibinfo{person}{Jessica Vitak}, \bibinfo{person}{Yuting
  Liao}, \bibinfo{person}{Mega Subramaniam}, {and} \bibinfo{person}{Priya
  Kumar}.} \bibinfo{year}{2018}\natexlab{}.
\newblock \showarticletitle{'I Knew It Was Too Good to Be True: The Challenges
  Economically Disadvantaged Internet Users Face in Assessing Trustworthiness,
  Avoiding Scams, and Developing Self-Efficacy Online}.
\newblock \bibinfo{journal}{\emph{Proceedings of the ACM on Human-Computer
  Interaction}} \bibinfo{volume}{2}, \bibinfo{number}{CSCW}
  (\bibinfo{year}{2018}), \bibinfo{pages}{176}.
\newblock


\end{thebibliography}

\end{document}